# Freezing in polyampholytes globules: Influence of the long-range nature of the interaction

Hindrik Jan Angerman and Eugene Shakhnovich
*Department of Chemistry and Chemical Biology, Harvard University, Cambridge, Massachusetts 02138, USA*

In random heteropolymer globules with short-range interactions between the monomers, freezing takes place at the microscopic length scale only, and can be described by a 1-step replica symmetry breaking. The fact that the long-range Coulomb interaction has no intrinsic length scale suggests that freezing in random polyampholyte globules might take place at all length scales, corresponding to an overlap parameter $q(x)$ that increases continuously from zero to its maximum value. Study of the polyampholyte globule within the independent interaction approximation seems to confirm this scenario. However, the independent interaction model has an important deficiency: it cannot account for self-screening, and we show that the model is only reliable at length scales shorter than the self-screening length. Using the more realistic sequence model we prove that in the general case of a random heteropolymer globule containing two types of monomers such that unlike monomers attract each other, freezing at arbitrarily large length scales is not possible. For polyampholyte globules this implies that beyond the self-screening length, the freezing behavior is qualitatively the same as in the case of short-range interactions. We find that if the polyampholyte globule is not maximally compact, the degree of frustration is insufficient to obtain freezing.

## I. INTRODUCTION

A polyampholyte is a polymer containing both positively and negatively charged monomers. Despite the fact that polyampholytes provide simple models for electrostatic interactions in biopolymers such as proteins, they have been studied intensively only during the last 10 years. Most attention has been paid to the conformations of isolated molecules in solution.[1-6] The authors of Ref. 1 considered a chain with zero net charge, containing a fraction $f$ of charged monomers randomly distributed along the chain. They showed that if the chain is sufficiently long, it collapses to form a globule (see also Ref. 7, where it was shown that oppositely charged polyelectrolyte chains merge to form globular complexes). Generally, the globular state[8,9] of a polymer chain is characterized by a well-defined, non-zero density, and by the fact that its correlations are short-range (short compared to the linear size of the globule). Thus, the globular state differs qualitatively from the coil-state. Depending on the value of $f$, this globule may or may not be maximally compact. In either case, further collapse is prevented by excluded volume interactions.[1] In order to find the density of non-compact polyampholyte globules, the authors of Ref. 1 used the Debye-Hückel theory[10] to describe the charge density fluctuations inside the globule. This is justified provided that the fraction $f$ of charged monomers is small, and the quenched sequence of charges is random.[2] Things might change if the polymer carries a net charge. In Refs. 3,4 it was shown via computer simulations that if each monomer carries a charge $\pm e$, then even a small net charge of order $e\sqrt{N}$ is sufficient to stretch short chains ($N$ is the number of monomers per chain). In Ref. 5 the effect of a net charge was investigated theoretically. It was shown that there exists a critical value $Q_* \propto \sqrt{N}$ for the net charge $Q_n$ such that polymers with $|Q_n| < Q_*$ collapse into spherical globules, whereas polymers with $|Q_n| > Q_*$ collapse into elongated globules. The temperature dependence of the characteristics of the globule was derived in Ref. 6 using a Flory-type theory. Multi-chain effects were studied in Ref. 11. It was shown that for neutral polyampholyte samples in solution the globules in the supernatant are spherical, since molecules with an opposite excess charge form neutral complexes (the same as what happens in mixtures of oppositely charged polyelectrolytes[7]).

The aim of this paper is to investigate the possibility of "freezing" in polyampholyte globules, and in particular we are interested in the influence of the long-range character of Coulomb interactions. The freezing-phenomenon has already been studied extensively for the situation where the interactions between the monomers are short-range.[12–18] First we give a brief explanation of the concept of freezing in compact heteropolymer systems. Consider a heteropolymer consisting of several different monomer species, which are distributed randomly along the chain. At low temperatures the (compact) system tries to reach a low-energy state. However, due to the constraints imposed by the chain connectivity, it is not possible to minimize all interaction energies at the same time. Therefore, the system is "frustrated," and has many local free energy minima. At low temperatures the ergodicity is broken: the phase space of the polymer globule is divided in many phases (physical states). As in spin glasses, these phases are not related to each other by symmetries of the Hamiltonian (see Ref. 19 for a review on spin glasses and replica theory). It was conjectured in Ref. 12 that to a first approximation the energy levels of a polymer globule are independent from each other, and that they have a Gaussian distribution. This would imply that a compact heteropolymer globule with short-range interactions is equivalent to the random energy model,[20,21] as far as the thermodynamics is concerned. This was shown to be true for a microscopic model in Ref. 14. The ergodicity breaking of the random energy model can be described by a 1-step replica symmetry breaking.[20] The physical meaning of a 1-step replica symmetry breaking for polymers is as follows. Consider $n \gg 1$ heteropolymer chains, all having exactly the same monomer sequence (such chains are called replicas). At conditions where ergodicity breaking occurs, the conformations of two replicas either coincide exactly up to the microscopic length scale (modulo rotations and translations), or they have no structural relationships at all.[14] It follows that when the ergodicity is broken, the relative monomer positions within one chain are more or less fixed. One describes the situation by saying that the chain is "frozen" in one conformation. Protein folding is a special example of heteropolymer freezing, where the energy of one of the frozen conformations, called the native one, is much lower than that of the others.[22] The applicability of the random energy model to describe ergodicity breaking in polymers is not unlimited. For instance, it was shown in Ref. 15 that for low space dimensions $d \leq 2$ the replica symmetry breaking is of a more complicated nature. The reason for this qualitative difference between globules in $d > 2$ and $d \leq 2$ space dimensions can be explained in a simple way. As an example for the case $d > 2$ consider a globule in 3 dimensions. Let $\vec{r}$ be a point visited by the chain, and consider the sphere with center $\vec{r}$ and radius $R$. The random walk describing the chain conformation will need roughly $R^2$ steps in order to leave this sphere, giving it an average density $\rho \propto 1/R$. Therefore, a lot of empty space is left, and the chain has to return many times in order to create a compact globular state. The chain returns because it is "reflected" at the boundary of the globule, and relatively short pieces of chain with length $\propto N^{2/3}$ traverse the whole volume. It follows that the chain is rather self-entangled, and so it is difficult to change the structure of part of the globule without affecting the structure of the remaining part. Therefore, two conformations that are different in some region of space will probably be different throughout the globule. This structural independence of different conformations makes that the random energy model provides a good description for $d = 3$. However, this reasoning breaks down for $d = 2$, because then the number of random walk steps needed to leave a circle with radius $R$ is proportional to the surface of the circle. In other words, if the polymer chain reaches the point $\vec{r}$, it will completely fill up the neighborhood of this point before moving on to other regions. This leads to a so-called *crumpled* globule, in which monomers that are close to each other in space are also close to each other along the chain. It follows that globules in $d \leq 2$ dimensions are always crumpled. Since the various parts of a crumpled globule do not interpenetrate and are therefore independent from each other, it is conceivable that the conformations of two crumpled globules coincide in some region of space, while they are completely different outside this region. Consequently, the assumption of structural independence of different conformations, the cornerstone of the random energy model, breaks down in low dimensional space. Another vital ingredient for the applicability of the random energy model is the short-range character of the interactions. If the interaction has a length scale $L$ larger than the microscopic length scale, then by performing a renormalization transformation[23] one can see that the random energy model is applicable on length scales $R \gtrsim L$, but it is not clear what happens at shorter length scales. If the interaction has no characteristic length scale at all (as it is the case for the Coulomb potential $V(r) \propto r^{2-d}$), the random energy model might not be applicable at any length scale. Application of the qualitative investigation of Section 4 in Ref. 15 to polyampholytes seems to confirm this, and suggests the existence of a critical dimension $d = 6$ below which the random energy model breaks down, implying a non-trivial replica symmetry breaking. However, as we will see, part of this conclusion is an artefact of the model employed (the independent interaction model). As suggested in Ref. 24 and confirmed by the present paper, at length scales larger than the polymer self-screening length the random energy assumption becomes valid, since the Coulomb interaction is effectively short-range at those length scales. However, it is worthwhile to investigate what is happening on smaller length scales. This might also be relevant for the problem of protein folding. Of the 20 natural occurring amino acids, two of them are positively charged, and two of them are negatively charged. At first sight it seems that Coulomb interactions are irrelevant because proteins fold under physiological conditions where the salt concentration is high and electrostatic interactions are effectively screened. However, screening can only take place in the coil-state of the chain. Once the protein is collapsed into a globule, all water and salt is squeezed out, and the electrostatic interactions between charged groups in the interior are unscreened.

The main part of this paper deals with freezing in the globular state of a weakly charged "statistically neutral" polyampholyte chain in a $\theta$-solvent or a poor solvent, because for these situations a rigorous analysis is possible. As described above, the globular state can arise in various situations. It can arise by the collapse of a single chain, leading to an elongated or spherical globule, depending on the net charge.[5] The globular state can also arise by the simultaneous collapse of two chains whose net charges cancel each other, leading to a spherical globule.[11] Finally, the globular state is present in the precipitate of a polyampholyte solution. Although in the latter two cases the system consists of several chains, the single-chain analysis of this paper remains valid provided that the chains are long enough. This is because even in the case of a single long globular chain, monomers that are far separated from each other along the chain (but that are possibly close to each other in space) do not "know" that they are attached to each other. This is due to the short correlation length inside the globule,[9] making the long loop connecting these monomers invisible.

Section II provides a summary of the model introduced in Ref. 1 to describe the globular state of a weakly charged polyampholyte. Section III presents the theoretical tools developed in Ref. 15 to analyze freezing in situations where the self-overlap of the replicas is small. In the case at hand this criterion is satisfied: if there is freezing, each replica will be frozen on the self-screening length scale, which is much larger than the microscopic length scale since the polyampholyte is only weakly charged. In Section IV we analyze the polyampholyte globule using the independent interaction model, leading to the prediction that in the absence of salt the overlap parameter[14] increases continuously from zero. In Section V we show that the independent interaction model provides a correct description only at length scales smaller than the self-screening length, and on larger length scales the solution found in section IV must be rejected. By using the more realistic sequence model[16–18] we prove that the overlap parameter cannot attain arbitrarily small values, and so $q(x)$ must be discontinuous even in the absence of salt. In Section VI, finally, we summarize our results, and discuss briefly the possibility of freezing in the globular state of highly charged polyampholytes.

## II. PHYSICAL MODEL

To describe the globular state of a weakly charged polyampholyte we follow Higgs and Joanny.[1] They considered an overall neutral random polyampholyte chain containing both charged and neutral monomers. Due to electrostatic interactions, the chain is collapsed to form a globule. Since the chain is weakly charged, the globule does not have maximum density, in other words, there is still solvent present inside the globule. We mainly consider the case where the solvent is either under $\theta$-conditions, or under poor-solvent conditions with respect to the neutral monomers. In both cases the chains obey random walk statistics on all length scales[9] in between the microscopic length scale and the linear size of the globule. This property is used explicitly in our calculations. If the solvent is good, then at length scales larger than the blob size (which equals[1] the polymer self-screening length) the chain statistics is again Gaussian, and on these length scales the behavior of the system is qualitatively the same as that under $\theta$-conditions. In this paper it will tacitly be assumed everywhere that the solvent is under $\theta$-conditions, unless stated otherwise. The composition of the chain is determined by the fraction $f \ll 1$ of charged monomers. Whether the number of neutral monomers in between two charged monomers is constant or random is of no importance. Apart from the sign, all charged monomers carry the same charge $\pm e$. The extrapolation of the results to multiply charged monomers is trivial. There is no charge correlation along the chain; in other words the sequence of charges is Bernoullian (random). The typical root-mean-square distance $a$ between two charged monomers neighboring along the chain is given by $a^2 \approx b^2/f$, where $b$ is the size of a neutral monomer. It was shown in Ref. 1 that in the globular state, further collapse is prevented by the excluded volume repulsion (the entropy loss due to confinement is a surface effect and can be neglected for long enough chains (volume approximation)). In case of non-compact globules the excluded volume repulsion is quantified by the third order virial coefficient in case of a $\theta$-solvent or a poor solvent, and by the second order virial coefficient in case of a good solvent. Under $\theta$-conditions the resulting globule density $\rho$, defined as the number of charged monomers per unit of volume, is given by[1]

$$\rho \approx \frac{f^2 \ell}{b^4} \qquad (1)$$

where $\ell = e^2/4\pi\varepsilon k_B T$ is the Bjerrum length; $\ell \approx 0.7 nm$ in water at room temperature. We require that $b > \ell$, because otherwise the charged monomers would stick together. In deriving Eq. (1) the authors of Ref. 1 assumed that the charge density fluctuations inside the globule can be described by the Debye-Hückel theory,[10] which is justified[2] if the polymer self-screening length $\kappa_p^{-1}$ is much larger than the average distance $\rho^{-1/3}$ between two charged monomers. Using Eq. (2) one finds that this condition is equivalent to $b^2 \gg f \ell^2$. At the same time, this implies that the globule density as given by Eq. (1) is much smaller than the maximum density $\rho_{max} = f/b^3$. The globular state can be considered as a melt of blobs.[25] Both under $\theta$-conditions, and under good solvent conditions the blob size coincides[1] with the polymer self-screening length $\kappa_p^{-1}$. The number of charged monomers $N_b$ per blob, and the radius $R_b$ of a blob are given by[1]

$$N_b \approx \frac{b^2}{f\ell^2} \gg 1 \qquad R_b \approx \kappa_p^{-1} \approx \frac{b^2}{f\ell} \qquad (2)$$

The melt of blobs is moderately strong interacting, which can be seen in the following way. The characteristic interaction energy $E$ between two neighboring blobs can be estimated as $E \approx Q_b^2/4\pi\varepsilon R_b \approx k_B T$, where $Q_b \approx e\sqrt{N_b}$ is the characteristic charge of a blob. Since the Coulomb interactions are screened at distances larger than $R_b$, on length scales $R \gg R_b$ the globule is equivalent to a (maximally compact) chain with length $N/N_b$ having random short-range interactions (the fact that this equivalence also holds with respect to the freezing behavior is one of the conclusions of this paper). In Ref. 16 it was shown that such a system has a freezing transition when the dispersive interaction energy between different monomers attains a critical value, which is of the order $k_B T$ (corresponding to $\chi_{\text{Flory}} \approx 1$). So, it may be concluded that the polyampholyte globule is at least not too far away from a freezing transition. This means that a considerable amount of frustration must be present in the system. The origin of this frustration can be understood in the following way. The physical state of a polyampholyte chain is dominated by conformations in which the charges screen each other (note that in *random* conformations the charges are unscreened). In order to achieve this so-called self-screening, the charges have to rearrange spatially. This rearrangement is hindered by the chain connectivity, leading to a frustrated situation.

## III. GAUSSIAN APPROXIMATION

On length scales smaller than the blob size, the thermal energy exceeds the electrostatic energy, and so freezing could only occur on length scales $R \gtrsim \kappa_p^{-1}$. Since it follows from $N_b \gg 1$ that $\kappa_p^{-1}$ is much larger than the microscopic length scale $a$, it is justified to use the procedure developed in Ref. 15 by Shakhnovich and Gutin. In this section we present a brief summary of their method. Within the framework of replica theory[19] the onset of freezing in a polymer globule is signaled by the appearance of a non-zero value for the overlap order parameter $Q_{\alpha\beta}(\vec{r},\vec{r}')$, which is for $\alpha \neq \beta$ defined by[14]

$$Q_{\alpha\beta}(\vec{r},\vec{r}') = \left\langle \sum_{i=1}^{N} \delta_a(\vec{r}-\vec{r}_\alpha^i) \, \delta_a(\vec{r}'-\vec{r}_\beta^j) \right\rangle \tag{3}$$

$\delta_a$ is the Dirac delta function smeared out over the microscopic distance $a$, the vector $\vec{r}_\alpha^i$ is the position of the $i^{\text{th}}$ (charged) monomer in replica $\alpha$, and the brackets denote a thermodynamic average. The order parameter $Q_{\alpha\beta}$ appears in a natural way during the calculation of the free energy, but it is convenient to define a simpler overlap parameter[14] $q_{\alpha\beta}$ having a clear physical interpretation:

$$q_{\alpha\beta} \equiv \frac{a^3}{N} \int d\vec{r} \, Q_{\alpha\beta}(\vec{r},\vec{r}) = \left\langle \frac{a^3}{N} \sum_{i=1}^{N} \delta_a(\vec{r}_\alpha^i - \vec{r}_\beta^i) \right\rangle \tag{4}$$

The quantity between brackets gives for replicas $\alpha$ and $\beta$ the fraction of monomers that occupy the same position in space (measured within accuracy $a$). The normalization $a^3/N$ makes $q_{\alpha\beta}$ dimensionless and ensures that $q_{\alpha\beta} = 1$ if replicas $\alpha$ and $\beta$ have perfect overlap up to the microscopic length scale $a$. Another important characteristic of the system is the density profile $\rho_\alpha(\vec{r})$, which is defined by

$$\rho_\alpha(r) = \left\langle \sum_{i=1}^{N} \delta_a(\vec{r} - \vec{r}_\alpha^i) \right\rangle \tag{5}$$

The density is in fact independent of $\alpha$. In deriving the equation for the free energy we will need an expression for the entropy $S_n = k_B \ln \Omega_n$, where $\Omega_n$ is the (weighted) volume in the phase space of $n$ ideal non-interacting chains, associated with a certain density profile $\rho_\alpha(\vec{r})$ and overlap parameter $Q_{\alpha\beta}(\vec{r},\vec{r}\,')$. Generalizing the method presented in Refs. 8,9,15 one can derive the following relations:

$$\frac{S_n}{k_B} = \int d\vec{x}_\alpha \, \varphi \, \hat{g}\varphi \, \ln \frac{\hat{g}\varphi}{\varphi}$$

$$Q_{\alpha\beta}(\vec{r},\vec{r}\,') = \int d\vec{x}_\alpha \, \delta(\vec{x}_\alpha - \vec{r}) \, \delta(\vec{x}_\beta - \vec{r}\,') \, \varphi \, \hat{g}\varphi \tag{6}$$

$$\rho_\alpha(\vec{r}) = \int d\vec{x}_\alpha \, \delta(\vec{x}_\alpha - \vec{r}) \, \varphi \, \hat{g}\varphi$$

where $d\vec{x}_\mu \equiv d\vec{x}_1 \cdots d\vec{x}_n$, and $\varphi(\vec{x}_\alpha)$ is some unknown function. The integral operator $\hat{g}$ is defined by[8]

$$\hat{g}\varphi(\vec{x}_\alpha) = \int d\vec{y}_\alpha \, \varphi(\vec{x}_\alpha) \prod_\alpha g(\vec{x}_\alpha - \vec{y}_\alpha) \tag{7}$$

$g(\vec{r}) = g(r)$ is the single bond distribution function; in other words, it is the probability density that in the ideal system, two (charged) monomers neighboring along the chain are separated by a distance $r$. Using the fact that the characteristic length scale $R$ of the function $\varphi$ (which is the scale at which freezing occurs) is much larger than the characteristic length scale $a$ of the function $g(r)$ (because $R \gtrsim \kappa_p^{-1} \gg a$), it is justified to make the approximation

$$\hat{g}\varphi \cong \varphi + \frac{a^2}{6} \sum_{\alpha,j} \frac{\partial^2 \varphi}{(\partial x_\alpha^j)^2} \tag{8}$$

where $j = 1,2,3$ denotes a spatial index. With this approximation, the relations in Eq. (6) become

$$\frac{S_n}{k_B} = \frac{a^2}{6} \int d\vec{x}_\mu \, \varphi \sum_{\alpha,j} \frac{\partial^2 \varphi}{(\partial x_\alpha^j)^2}$$

$$Q_{\alpha\beta}(\vec{r},\vec{r}\,') = \int d\vec{x}_\mu \, \delta(\vec{x}_\alpha - \vec{r}) \, \delta(\vec{x}_\beta - \vec{r}\,') \, \varphi^2 \tag{9}$$

$$\rho_\alpha(\vec{r}) = \int d\vec{x}_\mu \, \delta(\vec{x}_\alpha - \vec{r}) \, \varphi^2$$

In Section IV the free energy $F$ will be written as a functional of $\varphi(x_\alpha)$. In order to find the mean-field expression for the overlap order parameter $Q_{\alpha\beta}$, the free energy has to be maximized[29] with respect to $\varphi$. Following Ref. 15, we consider Gaussian trial functions of the form

$$\varphi(\{\vec{x}_\alpha\}) = \varphi_0 \exp(-\tfrac{1}{2} \sum_{\alpha,\beta} k_{\alpha\beta} \vec{x}_\alpha \cdot \vec{x}_\beta) \qquad \varphi_0^2 = \pi^{3(1-n)/2} \rho D^{3/2} \tag{10}$$

The matrix $k_{\alpha\beta}$ is of the Parisi-type;[26] the value $k_{\alpha\alpha}$ at the diagonal may be different from zero. The dot represents an inner product in 3 dimensional space. Because of translation invariance on length scales small compared to the linear size of the globule, $\varphi$ must be translation invariant, which is equivalent to the condition

$$\sum_\beta k_{\alpha\beta} = 0 \qquad \forall \alpha \tag{11}$$

The quantity $D$ appearing in Eq. (10) is the determinant of the matrix obtained[15] by deleting the $\alpha^{th}$ row and the $\alpha^{th}$ column of the matrix $k_{\alpha\beta}$; this determinant is independent of $\alpha$. After substitution of the trial function Eq. (10) for $\varphi$ into Eq. (9), one finds the following expressions for the density and the overlap parameter:

$$Q_{\alpha\beta}(\vec{r}_1,\vec{r}_2) = \frac{\rho}{(\pi D_{\alpha\beta})^{3/2}} \exp\left(-\frac{(\vec{r}_1-\vec{r}_2)^2}{D_{\alpha\beta}}\right) \quad \Leftrightarrow \quad Q_{\alpha\beta}(\vec{k}) = \rho \exp\left(-\frac{D_{\alpha\beta} k^2}{4}\right) \tag{12}$$

$$q_{\alpha\beta} = \frac{a^3}{(\pi D_{\alpha\beta})^{3/2}} \qquad \rho_\alpha(\vec{r}) = \rho$$

The Parisi matrix $D_{\alpha\beta}$ is the determinant of the matrix obtained from $k_{\alpha\beta}$ by deleting the $\alpha^{th}$ row, the $\alpha^{th}$ column, the $\beta^{th}$ row, and the $\beta^{th}$ column, divided by $D$.[15] The physical meaning of Eq. (12) is illustrated in Fig. 1: the conformations of replicas $\alpha$ and β coincide if they are measured with accuracy $R \approx \sqrt{D_{\alpha\beta}}$. One should realize that Eq. (12) represents a class of trial functions, and that the real overlap parameter might be different. According to Eq. (12) a weak overlap between two replicas $\alpha$ and $\beta$ is achieved by assuming that their conformations are confined within the same "tube" whose radius $R >> \kappa_p^{-1}$ is constant throughout the globule. In reality, however, it is possible that in part of the globule the conformations coincide down to the self-screening length, while in the remaining part they are completely different. In fact, this is probably what happens in low-dimensional $d \leq 2$ globules with short-range interactions. Nevertheless, even in this case the class Eq. (12) of Gaussian trial functions describes the situation qualitatively. Adopting this approximation, the expression for the entropy $S$ becomes[15]

$$\frac{S}{k_B} = \lim_{n \to 0} \frac{1}{n} \frac{S_n}{k_B} = -\frac{Na^2}{4} \int_0^1 \frac{dx}{x^2} K(x) \qquad K(x) = -xk(x) + \int_0^x dy\, k(y) \qquad (13)$$

The one-to-one correspondence between the function $k(x)$ and the off-diagonal elements of the Parisi matrix $k_{\alpha\beta}$ is defined in the usual way.[26] The function $K(x)$ is related to the function $D(x)$ (corresponding to the matrix $D_{\alpha\beta}$) via[15]

$$D(x) = 2\left(\frac{1}{xK(x)} - \int_x^1 \frac{dy}{y^2 K(y)}\right) \qquad (14)$$

## IV. INDEPENDENT INTERACTION MODEL

In this section we study the weakly charged polyampholyte globule in the independent interaction approximation,[14,15] which is based on the simplifying assumption that the coupling constants $\{B_{ij}\}$, determining the interaction strength between charged monomers $i$ and $j$, are independent from each other. This assumption is violated in the case of interacting electric charges, where $B_{ij}$ is proportional to the product of the charges carried by monomers $i$ and $j$. For instance, if $B_{ij} < 0$ and $B_{ik} < 0$, then necessarily $B_{jk} > 0$. However, our primary interest is to find the influence of the long-range nature of the Coulomb interaction, and the model is capable of capturing this feature. In Section V we will drop the independent interaction assumption by studying the more realistic sequence model. The great advantage of the independent interaction model with respect to the sequence model is, that it is technically easier to handle, which is the reason why it was considered in the earlier theoretical papers concerned with freezing in copolymers with a short-range interaction.[14,15] We will see that, within the Gaussian variational approach used throughout this article, the independent interaction model can be solved analytically. However, in case of Coulomb interactions a very important drawback of the model is that it is not capable of describing screening, which can be seen in the following way. Consider any monomer, say monomer $i$. This monomer tends to be surrounded by a cloud of monomers $k$ for which $B_{ik} < 0$. Now suppose that monomer $i$ is approached by a repelling monomer $j$ (that is, $B_{ij} > 0$). In the physical system of a mixture of positive and negative charges, monomer $j$ is attracted by the cloud, thus the unfavorable interaction between $i$ and $j$ is screened. If, on the other hand, the interaction strengths are independent, then the fact that $i$ repels $j$, whereas $i$ attracts $k$, does not give any information about the interaction between $j$ and $k$. Therefore, as experienced by $j$, the cloud is just a random collection of charges and the unfavorable interaction between $i$ and $j$ is unscreened. This suggests that the independent interaction model can only describe the polyampholyte globule correctly on length scales that are smaller than the polymer self-screening length $\kappa_p^{-1}$, which is confirmed by the analysis in Section V. Since it will turn out that on scales $R \gg \kappa_p^{-1}$ the system is qualitatively equivalent to a globule with short-range interactions, the independent interaction model describes the most promising regime for finding non-trivial replica symmetry breaking. On the other hand, at length scales smaller than $\kappa_p^{-1}$ the thermal energy becomes of the same order of magnitude as the electrostatic energy, so *if* there is freezing at a small length scale it has to be weak (see also the discussion at the end of Section II).

We consider the physical system described in Section II, and model it using the independent interaction approximation. We will assume a Gaussian distribution for the interaction strengths $B_{ij}$. At first sight, this choice seems difficult to justify, because after renormalization, $B_{ij}$ is the product of two variables with a Gaussian distribution (these variables represent the charges of the coarse grained monomers), which is itself not Gaussian. However, we note that expansion of the sequence model in the regime $R \ll \kappa_p^{-1}$ leads to the same expression for the free energy as the independent interaction model with a Gaussian distribution for $B_{ij}$ (compare Eq. (23), which is valid for the independent interaction model, with Eq. (48), which is derived from the sequence model). In the continuous representation of the chain,[27] where the parameter $t$ "numbers" the charges, the partition function can be written as

$$Z = \int d\vec{r}(t) \exp\left[-\frac{3}{2a^2} \int dt \left(\frac{d\vec{r}(t)}{dt}\right)^2 - \frac{\ell}{2} \int dt\, dt'\, B(t,t') V(r(t), r(t'))\right] \quad (15)$$

In writing down Eq. (15) it was tacitly assumed that the integration over chain conformations $\vec{r}(t)$ is restricted to conformations having a constant density $\rho(\vec{r}) \equiv \rho$ as defined in Eq. (5) (see the discussion following Eq. (41) for a justification of our neglect of density fluctuations). The Gaussian probability distribution for $B(t,t')$ has mean zero and variance unity.

$$P(B(t,t')) = \frac{1}{\sqrt{2\pi}} \exp\left[-\tfrac{1}{2} B^2(t,t')\right] \tag{16}$$

For the potential $V$ we take the screened Coulomb potential[10]

$$V(\vec{r},\vec{r}') = V(|\vec{r}-\vec{r}'|) \qquad V(r) = \frac{\exp[-\kappa r]}{r} \tag{17}$$

Within the limitations of the independent interaction model, $\kappa^{-1}$ represents the screening length due to the presence of salt, in which case it is an independent parameter of the theory. We will see, however, that this physical interpretation for $\kappa^{-1}$ leads to qualitatively wrong results. When the independent interaction model is obtained from the sequence model by expansion of the free energy in powers of the potential (see Section V), $\kappa^{-1}$ turns out to be the self-screening length $\kappa_p^{-1}$ of the polymer globule, which means that it can not be considered as an independent parameter. This fact is closely related to the observation that charges interacting via independent interactions cannot screen each other. Nevertheless, it is instructive to analyze the independent interaction model for general values of $\kappa^{-1}$. Following the same kind of derivation as presented in Refs. 16,17 one arrives at the following expression for the partition function $[Z^n]$ of $n$ replicas averaged over all sequences:

$$[Z^n] \propto \int dQ_{\alpha\beta}(\vec{r},\vec{r}') \, \exp\left[\frac{S_n(Q_{\alpha\beta})}{k_B} + \frac{\ell^2}{8} \sum_{\alpha\neq\beta} \int d\vec{r}_1 d\vec{r}_2 d\vec{r}_3 d\vec{r}_4 \, Q_{\alpha\beta}(\vec{r}_1,\vec{r}_2) \, Q_{\alpha\beta}(\vec{r}_3,\vec{r}_4) \, V(\vec{r}_1,\vec{r}_3) \, V(\vec{r}_2,\vec{r}_4)\right]$$

$$S_n(Q_{\alpha\beta}) \equiv k_B \ln \int d\vec{r}(t) \exp\left[-\frac{3}{2a^2} \sum_\alpha \int dt \left(\frac{d\vec{r}_\alpha(t)}{dt}\right)^2\right] \delta[Q_{\alpha\beta}(\vec{r},\vec{r}') - \hat{Q}_{\alpha\beta}(\vec{r},\vec{r}')] \, \delta[\rho - \hat{\rho}_\alpha(\vec{r})]$$

$$\tag{18}$$

The expression for the entropy $S_n$ has been given in Section III, Eqs. (9) and (13). Factors that do not depend on the overlap parameter $Q_{\alpha\beta}$ were omitted from Eq. (18). The hats on the functions $\hat{Q}_{\alpha\beta}(\vec{r},\vec{r}')$ and $\hat{\rho}_\alpha(\vec{r})$ indicate that they are functions of the microscopic state $\{\vec{r}_\alpha(t)\}$ of the system. They are defined as in Eqs. (3) and (5), but without averaging: $\hat{Q}_{\alpha\beta}(\vec{r},\vec{r}') = \int dt \, \delta(\vec{r}-\vec{r}_\alpha(t)) \delta(\vec{r}'-\vec{r}_\beta(t))$, and $\hat{\rho}_\alpha(\vec{r}) = \int dt \, \delta(\vec{r}-\vec{r}_\alpha(t))$. Note that $Q_{\alpha\beta}(\vec{r},\vec{r}')$ appearing in Eq. (18) is just a dummy integration variable. It is convenient to write Eq. (18) in the following form:

$$[Z^n] = \int dQ_{\alpha\beta}(\vec{r},\vec{r}') \, \exp\left[\frac{S_n}{k_B} - \frac{E_n}{k_B T}\right] \equiv \int dQ_{\alpha\beta}(\vec{r},\vec{r}') \, \exp[-\frac{H_n^{\text{eff}}}{k_B T}] \tag{19}$$

The definition for $E_n$ is clear from comparison with Eq. (18):

$$\frac{E_n}{k_B T} = -\frac{\ell^2}{8} \sum_{\alpha\beta} \int d\vec{r}_1 d\vec{r}_2 d\vec{r}_3 d\vec{r}_4 \, Q_{\alpha\beta}(\vec{r}_1, \vec{r}_2) Q_{\alpha\beta}(\vec{r}_3, \vec{r}_4) V(\vec{r}_1, \vec{r}_3) V(\vec{r}_2, \vec{r}_4) \qquad (20)$$

In the following we will refer to $S_n$ as the "entropy," and to $E_n$ as the "energy," although strictly speaking this is not correct. The great advantage of the independent interaction model as compared to the sequence model is the relatively simple expression for the energy. We will first present a highly simplified qualitative analysis that will reveal the existence of a critical value $d_c = 6$ for the dimension $d$. We will show that for $d > d_c$ freezing can only take place at the microscopic length scale, implying the validity of the random energy assumption, while for $d < d_c$ the freezing is of a more complicated nature. Following Ref. 15 we assume that $\varphi$ has a characteristic scale $R \gg a$, and can, therefore, be written as

$$\varphi(\vec{x}_1, \cdots, \vec{x}_n) = R^{d(1-n)/2} \varphi_1\left(\frac{\vec{x}_1}{R}, \cdots, \frac{\vec{x}_{n-1}}{R}\right) \qquad (21)$$

Note that due to translation invariance the function $\varphi_1$, which has unit scale and unit value, can be written as a function of $(n-1)$ arguments. The exponent of the scaling relation follows from the normalization $\int dx \, \varphi^2(x) = N$ of $\varphi$ (this normalization condition can be found by integrating the density $\rho_\alpha(\vec{r})$ given in Eq. (9) over the volume of the system). The idea is to find the scaling relations for the entropy and the energy, and to see whether the free energy has a stable extreme. The scaling relation for the entropy was derived in Ref. 15 to be $S \propto R^{-2}$. The scaling relation for the energy Eq. (20) can be found by inserting for $V(\vec{r})$ the generalized Coulomb potential $V(\vec{r}) \propto r^{2-d}$, for the overlap parameter $Q_{\alpha\beta}(\vec{x}, \vec{y})$ the expression given in Eq. (9), and for $\varphi(\vec{x}_1, \cdots, \vec{x}_n)$ the scaling relation Eq. (21). After some work one finds eventually $E \propto r^{-d+4}$, and the free energy becomes

$$F \approx \frac{A_1}{R^2} - \frac{A_2}{R^{d-4}} \qquad (22)$$

For $n < 1$ both constants $A_1$ and $A_2$ are negative,[15] and maximization of the free energy will yield a non-trivial value for $R$ if $d < 6$ (note that for short-range interactions the critical dimension is given by $d_c = 2$). In order to improve upon this rough analysis, it is necessary to use the more sophisticated trial function for $\varphi$ that has been given in Eq. (10). First we rewrite expression Eq. (20) for the energy, using the translation invariance inside the polymer globule.

$$\frac{E_n}{k_B T} = -\frac{N\ell^2}{8\rho} \frac{1}{(2\pi)^3} \int d\vec{k}\, V^2(k) \sum_{\alpha \neq \beta} Q_{\alpha\beta}^2(k) \tag{23}$$

Inserting the trial function Eq. (12) for $Q_{\alpha\beta}$, and taking for $V$ the screened Coulomb potential Eq. (17), one obtains

$$\frac{E_n}{k_B T} = -N\ell^2 \rho \int_0^\infty dk\, \frac{k^2 \exp[-\tfrac{1}{2} D_{\alpha\beta} k^2]}{(\kappa^2 + k^2)^2} \tag{24}$$

After taking the limit $n \to 0$ the energy becomes

$$\frac{E}{k_B T} = \lim_{n \to 0} \frac{1}{n} \frac{E_n}{k_B T} = N\rho \kappa^{-1} \ell^2 \sqrt{\pi/8} \int_0^1 dx\, f\!\left(\kappa^2 D(x)\right)$$

$$f(t) = -\sqrt{t} + \sqrt{\pi/2}\,(1+t)\,\exp[t/2]\left(1 - \Phi\!\left(\sqrt{t/2}\right)\right) \tag{25}$$

$$\Phi(x) = \frac{2}{\sqrt{\pi}} \int_0^x dy\, \exp[-y^2]$$

$\Phi$ is the error function. For future reference it is useful to give the series expansion and the asymptotic expansion[28] for $f$

$$\begin{aligned} f(t) &\cong \sqrt{\pi/2} - 2t^{1/2} \quad &\text{for} \quad t \ll 1 \\ f(t) &\cong 2t^{-3/2} \quad &\text{for} \quad t \gg 1 \end{aligned} \tag{26}$$

Using the asymptotic expansion, an interesting observation can be made. It follows that for $\kappa^2 D \gg 1$ the energy can be approximated by

$$E \propto \int_0^1 dx\, D^{-3/2}(x) \qquad \text{for} \quad \sqrt{D} \gg \kappa^{-1} \tag{27}$$

which coincides with the expression for the energy derived in Ref. 15 for the case of short-range interactions. Since according to Eq. (12) $\sqrt{D_{\alpha\beta}}$ is the freezing length scale (by which we mean the length scale at which the conformations of replicas $\alpha$ and $\beta$ coincide), one can conclude that with respect to the freezing behavior the interaction is effectively short-range on length scales $R \gg \kappa^{-1}$. Taking Eq. (13) for the entropy $S$, and ignoring numerical factors of order unity, the final expression for the free energy $F$ in the independent interaction approximation becomes

$$\frac{\beta F}{N\kappa^2 a^2} = -\int_0^1 \frac{dx}{x^2} \tilde{K}(x) + c^3 \int_0^1 dx\, f(\tilde{D}(x)) \tag{28}$$

$$c = \kappa_p/\kappa \qquad \tilde{K}(x) = \kappa^{-2} K(x) \qquad \tilde{D}(x) = \kappa^2 D(x)$$

Note that the dimensionless quantities $\tilde{K}$ and $\tilde{D}$ are still related via Eq. (14). If the independent interaction model is derived from the sequence model, then $c$ is of order unity (Section V). However, we will analyze the model for general values of $c$, and find that non-trivial replica symmetry breaking occurs for $c \gtrsim 1$. The free energy $F$ must be maximized,[29] so a necessary condition for $\tilde{K}$ is obtained by differentiating Eq. (28) with respect to $\tilde{K}(x)$, using Eq. (14). By differentiating the result with respect to $x$, one obtains

$$\tilde{D}(x) = (f^{(2)})^{-1}\left(\frac{\tilde{K}^3(x)}{2c}\right) \qquad \text{or} \qquad \frac{d\tilde{K}}{dx} = 0 \tag{29}$$

where $f^{(2)}$ stands for the second derivative of $f$. Differentiating again with respect to $x$ leads to the following solution to the maximization condition of Eq. (28):

$$f^{(3)}\left[(f^{(2)})^{-1}\left(\frac{\mathsf{K}^3}{2}\right)\right] = -\tfrac{3}{4}\mathsf{K}^4 \mathsf{x} \qquad \text{or} \qquad \frac{d\mathsf{K}}{d\mathsf{x}} = 0 \tag{30}$$

The reduced quantities $\mathsf{x}$ and $\mathsf{K}$ are defined by

$$\mathsf{x} = cx \qquad \mathsf{K} = \frac{\tilde{K}}{c} = \frac{\kappa^{-2} K}{c} \tag{31}$$

The non-trivial solution given in Eq. (30) is shown graphically in Fig. 2. It exists only for $\mathsf{x} \geq \mathsf{x}_c \cong 1.49$, and it has two branches. Since the function $q(x)$ corresponding to the overlap matrix $q_{\alpha\beta}$ given in Eq. (12) has to be an increasing function of $x$, it follows that the upper branch is the physical one. The asymptotic behavior of $\mathsf{K}$ for large $\mathsf{x}$ can be found using the series expansion of $f$, leading to

$$\mathsf{K} \cong \mathsf{x} \qquad \text{for} \quad \mathsf{x} \to \infty \tag{32}$$

In terms of the original variable $x$, this solution exists on the interval $x_c/c \leq x \leq 1$, provided that $c > x_c$. However, one can check that the two expressions Eqs. (14) and (29) for $D(x)$ do not coincide. Therefore, there must[30] be an interval $x_0 \leq x \leq 1$ in which the second solution given in Eq. (30) is valid, that is, $K$ is constant on this interval. The value of this constant $K_0$ is, given $x_c/c < x_0 < 1$, fixed by Eqs. (14) and (29). The maximization condition for the free energy, together with the physical condition that $K(x)$ must be a non-decreasing function of $x$, imply that $K(x)$ is continuous at $x = x_0$, which fixes both $x_0$ and $K_0$. It follows that $\tilde{K}_0 \equiv \kappa^{-2} K_0$ must be a root of the following equation:

$$f^{(2)}(2/\tilde{K}) = \tilde{K}^3/2c^3 \tag{33}$$

Eq. (26) implies that the left hand side of Eq. (33) is proportional to $\tilde{K}^{3+\frac{1}{2}}$ for small $\tilde{K}$, while it is proportional to $\tilde{K}^{3/2}$ for large $\tilde{K}$. This means that for small $c$ Eq. (33) has no solution at all, while for large $c$ it has two solutions, the larger of the two being the physical one. Using the series expansion for $f$, we obtain for $c \gg 1$

$$\tilde{K}_0 \cong c^2/2 \tag{34}$$

Under the same condition $c \gg 1$, the non-constant part of the solution follows from the combination of Eqs. (31) and (32)

$$K(x) \cong c^2 x \qquad x \geq x_c \propto c^{-1} \tag{35}$$

The value for $x_0$ follows from continuity at $x = x_0$:

$$x_0 \cong 1/2 \tag{36}$$

The plateau value $D_0$ for $D$ is given by (use Eq. (14))

$$D_0 = \frac{2}{K_0} = \frac{2}{\kappa^{-2}\tilde{K}_0} \cong \frac{4}{\kappa^{-2} c^2} \approx \kappa_p^{-2} \tag{37}$$

Since $x = 1$ corresponds to self-overlap, it follows from Eqs. (12) and (37) that each replica is frozen on the self-screening length scale. The presence of a plateau means that there is a non-zero probability that two different replicas have perfect overlap down to the self-screening length scale. The non-constant part of $K(x)$, which is absent in the case of short-range interactions, corresponds to the possibility that the conformations of two replicas coincide down to length scale $R$, with $\kappa_p^{-1} < R < \kappa^{-1}$. The existence of this interval for $R$ is physically reasonable, because at length scales shorter than $\kappa_p^{-1}$ the thermal energy $k_B T$ is larger than the electrostatic energy, while at length scales larger than $\kappa^{-1}$ the interaction is effectively short-range due to the screening. The resulting shape of $K(x)$ is shown in Fig. 3. Note that in the limiting case $\kappa = 0$ the overlap parameter $q(x)$ increases continuously from zero, meaning that there are replica pairs having an arbitrarily small overlap. This implies that there is no upper bound for the "tube diameter" $D_{\alpha\beta}^{-1}$ (see Fig. 1), and one could describe the situation by saying that the freezing occurs on all length scales larger than the self-screening length.

Let the overlap $q_{\alpha\beta}$ between two replicas $\alpha$ and $\beta$ be defined as in Eq. (12). Let $P(q)$ denote the probability density that two randomly picked replicas have overlap $q$. Then $P(q)$ can be obtained[31] from the inverse of $q(x)$ via $P(q) = dx/dq$. The dependence of $P(q)$ on $q$ is shown in Fig. 4. It has two delta peaks, one corresponding to the non-zero probability to find two replicas that have perfect overlap on the length scale $\kappa_p^{-1}$, the other corresponding to the probability that two replicas have no overlap at all. The continuous part, which is due to the long-range character of the interaction, shows that there is a multitude of weakly overlapping states (compare with Ref. 15).

## V. SEQUENCE MODEL

As explained in the previous section, self-screening does not exist in a system of particles interacting via independent coupling constants $B_{ij}$. Since at length scales $R > \kappa_p^{-1}$ self-screening is very important in the polyampholyte globule, it is necessary to use a more realistic model in order to find the physical relevance of the results obtained in Section IV. In the sequence model the composition of the chain is described by its monomer sequence $\vartheta_i$. The parameter $i = 1, \cdots, N$ counts the charged monomers in the order determined by the chain connectivity. If $\vartheta_i = +1$ ($\vartheta_i = -1$) then the $i^{\text{th}}$ charged monomer has a positive (negative) electric charge of magnitude $e$. For computational convenience it will be assumed that $\vartheta_i$ has a Gaussian distribution with mean zero and variance unity, i.e. the parameters $\vartheta_i$ are drawn independently from the distribution

$$P(\vartheta) = \frac{1}{\sqrt{2\pi}} \exp\left[-\tfrac{1}{2}\vartheta^2\right] \tag{38}$$

This simplification can be justified in the following way. On physical grounds, freezing might only occur at length scales $R \gtrsim \kappa_p^{-1}$, which is confirmed by the analysis in Section IV. Since $a \ll \kappa_p^{-1}$, it is possible to perform a renormalization transformation by defining coarse grained monomers each consisting of $N_1$ monomers, where $1 \ll N_1 \ll N_b$, without changing the qualitative behavior of the model. The coarse grained monomers have a Gaussian charge distribution, which justifies the above-mentioned simplification. In the continuous representation of the chain[27] the partition function takes on the form

$$Z = \int d\vec{r}(t) \exp\left[-\frac{3}{2a^2}\int dt \left(\frac{d\vec{r}(t)}{dt}\right)^2 - \frac{\ell}{2}\int dt\, dt'\, \vartheta(t)\vartheta(t') V(\vec{r}(t),\vec{r}(t'))\right] \quad (39)$$

Again, it is assumed implicitly that the integration over chain conformations is restricted to conformations for which the density as defined in Eq. (5) is constant. For the potential $V(\vec{r},\vec{r}') = V(|\vec{r}-\vec{r}'|)$ we substitute the unscreened Coulomb potential $V(r)=1/r$. In order to calculate the free energy using replica theory[19] one has to average the partition function $Z^n$ of $n$ non-interacting replicas over all possible sequences $\vartheta(t)$ weighted according to the distribution given in Eq. (38). Following Refs. 16–18, it is convenient to introduce the parameter $\hat{\psi}(\vec{r})$ describing the charge density fluctuations:

$$\hat{\psi}(\vec{r}) = \int dt\, \vartheta(t)\, \delta_a(\vec{r}-\vec{r}(t)) \quad (40)$$

It is defined as the deviation of the charge density from its average value (which is zero in our case). Since opposite charges attract each other (in Ref. 16 this situation was referred to as "the mixing case") the thermodynamic average of this parameter will always be zero. After writing the partition function in terms of $\hat{\psi}(\vec{r})$, it is possible to make the effective Hamiltonian linear in $\vartheta$ by performing the Hubbard-Stratonovich trick, enabling the calculation of the average over the sequences. One obtains (the entropy $S_n$ has been defined in Eq. (18); see also Refs. 16,17 for additional details)

$$[Z^n] = \int d\phi_\alpha(\vec{r}) \int d\psi_\alpha(\vec{r}) \exp\left[-\frac{\ell}{2}\sum_\alpha \int d\vec{r}d\vec{r}'\, V(\vec{r},\vec{r}')\psi_\alpha(\vec{r})\psi_\alpha(\vec{r}') + i\sum_\alpha \int d\vec{r}\, \phi_\alpha(\vec{r})\psi_\alpha(\vec{r}')\right] *$$

$$* \int dQ_{\alpha\beta}(\vec{r},\vec{r}') \exp\left[\frac{S_n(Q_{\alpha\beta})}{k_B} - \frac{1}{2}\sum_{\alpha\neq\beta}\int d\vec{r}d\vec{r}'\, Q_{\alpha\beta}(\vec{r},\vec{r}')\phi_\alpha(\vec{r})\phi_\beta(\vec{r}') - \frac{\rho}{2}\sum_\alpha \int d\vec{r}\, \phi_\alpha^2(\vec{r})\right]$$
(41)

This is a suitable moment to discuss the validity of the various approximations made in arriving at Eq. (41). One of the approximations is the assumption of a Gaussian probability distribution for the monomer charges, which causes the absence of terms of higher order in $\phi_\alpha(\vec{r})$. If the partition function $[Z^n]$ is calculated using the discrete distribution $P(\vartheta=+1) = P(\vartheta=-1) = 1/2$, fourth and higher order terms are generated. As discussed earlier, the use of a Gaussian probability distribution can be justified by considering a coarse graining transformation. The higher order terms missed in this way are unimportant due to the fact that the thermodynamic average of $\hat{\psi}_\alpha(\vec{r})$ is zero. Another approximation is the neglect of the fluctuations in the total density $\rho(\vec{r}) = \rho_+(\vec{r}) + \rho_-(\vec{r})$, where $\rho_+(\vec{r})$ and $\rho_-(\vec{r})$ stand for the local density of positively, resp. negatively charged monomers. Note that it would not be easy to get rid of this approximation, since the trial function Eq. (10) does not incorporate density fluctuations. At first sight one may wonder whether it is consistent to take the fluctuations in the charge density $\psi(\vec{r}) \equiv \rho_+(\vec{r}) - \rho_-(\vec{r})$ into account, while neglecting the fluctuations in the total density. However, the charge density and the total density play completely different roles. The fluctuations $\delta\psi$ in the charge density are responsible for the electrostatic attraction[10] (note that $\delta\psi = \psi$); in other words the energy of a globule in the coarse grained state $\{\psi(\vec{r}), \rho(\vec{r})\}$ is a function of $\delta\psi(\vec{r})$ alone, and does not depend on $\delta\rho(\vec{r})$ (see the first contribution to the effective Hamiltonian in Eq. (41)). Since this electrostatic interaction is responsible for the freezing transition, it is clear that the fluctuations $\delta\psi(\vec{r})$ in the charge density are essential, and must be taken into account even in the lowest order approximation. The fluctuations $\delta\rho(\vec{r})$ in the total density, on the other hand, are far less important. As can be seen from Eq. (41), they appear only in the combination $(\rho + \delta\rho)$, for instance as argument of the entropy $S_n(Q_{\alpha\beta}, \rho_\alpha)$. At the relevant length scales $R > \kappa_p^{-1}$, the fluctuations $\delta\rho$ are small compared to the average density $\rho$, and in the lowest order approximation it is justified to neglect them. Returning to Eq. (41), it is straightforward to proceed by calculating the Gaussian integrals over $\psi_\alpha(\vec{r})$ and $\phi_\alpha(\vec{r})$. The result is[16]

$$[Z^n] \propto \int dQ_{\alpha\beta}(\vec{r}, \vec{r}') \exp\left[-\frac{H_n^{\text{eff}}}{k_B T}\right] \qquad H_n^{\text{eff}} = E_n - TS_n$$

(42)

$$\frac{E_n}{k_B T} = \frac{1}{2} \frac{V}{(2\pi)^3} \int d\vec{k} \ln \det A_{\alpha\beta}(k)$$

The elements of the Parisi matrix $A$ are given by

$$\begin{cases} A_{\alpha\beta}(k) = \frac{1}{2} Q_{\alpha\beta}(k) & \alpha \neq \beta \\ A_{\alpha\alpha}(k) = \frac{1}{2}\rho + \frac{1}{2\ell V(k)} \end{cases}$$

(43)

Factors that are independent of the overlap parameter were omitted from Eq. (42). Since the chain is in a globular state, which is a well-defined state with short-range fluctuations,[8,9] it is reasonable to assume that the saddle point method[32] gives a good approximation to the integral over $Q_{\alpha\beta}(k)$. Consequently, the equilibrium value of the overlap parameter can be obtained by finding a stable extreme of the effective Hamiltonian $H_n^{\text{eff}}$. In the limit $n \to 0$ a stable extreme corresponds to a local maximum[29] of the mean-field free energy $F$ given by

$$F = \lim_{n \to 0} \frac{1}{n} H_n^{\text{eff}} \tag{44}$$

In order to calculate $F$, note that the expression for the entropy $S$ is the same as it was for the independent interaction model; see Eq. (13). To find the expression for the energy, we use the following identity[33] for Parisi matrices $A$

$$\frac{E(k)}{k_B T} \equiv \lim_{n \to 0} \frac{1}{n} \ln \det A_{\alpha\beta}(k) = \ln(\tilde{a} - \langle a \rangle) + \frac{a(0)}{\tilde{a} - \langle a \rangle} - \int_0^1 \frac{dx}{x^2} \ln\left(1 - \frac{[a](x)}{\tilde{a} - \langle a \rangle}\right) \tag{45}$$

$$\tilde{a} = A_{\alpha\alpha} \qquad \langle a \rangle = \int_0^1 dx\, a(x) \qquad [a](x) = x\, a(x) - \int_0^x dy\, a(y)$$

The quantities $\tilde{a}$, $\langle a \rangle$, and $[a](x)$ all depend on $k$. It is convenient to divide all matrix elements $A_{\alpha\beta}$ by $\rho/2$; from Eqs. (42) and (45) it follows that this corresponds to adding a constant to the free energy. Combining Eqs. (12), (43) and (45), and substituting for $V(r)$ the unscreened Coulomb potential $V(r) = 1/r$, one arrives at

$$\tilde{a} = 1 + \frac{1}{\rho \ell V(k)} = 1 + \kappa_p^{-2} k^2 \qquad a(x) = \exp\left[-\tfrac{1}{4} D(x) k^2\right]$$

$$\kappa_p^2 = 4\pi\rho\ell \tag{46}$$

where $\kappa_p^{-1}$ is the polymer self-screening length (note that $\rho$ differs by a factor 2 from the charge density $n$ used in Refs. 1,10). It was mentioned in Section IV that if the expression for the energy in the sequence model is expanded in powers of the potential, then the first term of this expansion gives the expression for the energy in the independent interaction model. Since Eq. (45) for the energy depends on the potential $V(k)$ via the parameter $\tilde{a}$, an expansion in powers of the potential is equivalent to an expansion in powers of $1/\tilde{a}$. This expansion is justified for $\vec{k}$-vectors with length $k \gg \kappa_p^{-1}$ (note that since $D(x)$ is positive, we have the inequalities $0 < \langle a \rangle < 1$, and $0 < \alpha(x) < 1$); corresponding to length scales shorter than the self-screening length. Under these conditions the energy Eq. (45) can be approximated by

$$\frac{E(k)}{k_B T} \cong \ln \tilde{a} + \frac{1}{2\tilde{a}^2} \int_0^1 dx\, a^2(x) \qquad \text{if } \tilde{a} \gg 1 \tag{47}$$

After neglecting terms that are independent of the overlap parameter, this expression can be rewritten as

$$\frac{E}{k_B T} \cong \frac{1}{2} \frac{V}{(2\pi)^3} \int d\vec{k} \frac{1}{2(1+\kappa_p^{-2} k^2)^2} \lim_{n \to 0} \frac{-1}{n} \sum_{\alpha \neq \beta} \left( \rho^{-1} Q_{\alpha\beta}(k) \right)^2 =$$
$$= -\frac{\ell^2}{4} \frac{V}{(2\pi)^3} \int d\vec{k} \, V^2(k; \kappa_p) \lim_{n \to 0} \frac{1}{n} \sum_{\alpha \neq \beta} Q_{\alpha\beta}^2(k) \tag{48}$$

where $V(k; \kappa_p) \equiv 4\pi / (\kappa_p^2 + k^2)$ is the Fourier transform of the screened Coulomb interaction with inverse screening length $\kappa_p$. As mentioned, this expression for the energy coincides with the one derived from the independent interaction model (see Eq. (23)), with the important difference that the parameter $\kappa^{-1}$ does not stand for the salt screening length, but for the polymer self-screening length. Therefore, $\kappa^{-1}$ is not a free parameter: the globule density and the temperature fix its value. Following the same lines as in Section IV, one arrives at Eq. (28) for the free energy with the parameter $c$ given by

$$c^3 = \frac{1}{2^{3/2} \pi} \frac{\ell^{1/2}}{\rho^{1/2} a^2} \tag{49}$$

If one inserts the expression for $\rho$ as given in Eq. (1), one finds $c \cong 0.5$. According to Section IV there exists a critical value $c_c \cong 1.5$ for $c$ such that for $c < c_c$ the non-trivial solution given in Eq. (30) does not exist (it can be shown that the critical value $c'_c$ for the existence of a 1-step replica symmetry breaking solution has roughly the same value). Of course, the expression $\rho \approx f^2 \ell / b^4$ for the density is very crude: it is based on hand waving arguments and all numerical factors of order unity have been omitted. Nevertheless, due to the very weak dependence $c \propto \rho^{-1/6}$ of $c$ on $\rho$ we conclude that the present analysis predicts the absence of replica symmetry breaking in a weakly charged polyampholyte globule under $\theta$-solvent conditions. Under poor-solvent conditions the situation is even worse since the density appears in the *denominator* of $c$. At first sight, this seems a bit contradictory: under poor solvent conditions the density is higher, and the electrostatic interactions are stronger. However, since the monomers are confined to a smaller volume, the constraints imposed by the polymer bonds become less important (the monomers are "asymptotically free"). They can rearrange themselves more easily in order to avoid unfavorable interactions, and the frustration is less than under $\theta$-conditions. This reasoning can be made more quantitative by noting that under poor-solvent conditions the blob size $R_b$ is smaller than the self-screening length $\kappa_p^{-1}$: their ratio scales with the density $\rho$ according to $R_b / \kappa_p^{-1} \propto \rho^{-1/2}$.

The expansion of $E(k)$, leading to the independent interaction approximation to the energy, is justified only for $k \gg \kappa_p^{-1}$, which corresponds to length scales shorter than the polymer self-screening length. In order to study the freezing behavior on larger length scales, one must resort to the full energy expression given by Eqs. (42), (43) and (45). By writing the entropy in terms of $D(x)$ using

$$\frac{2}{K(x)} = xD(x) + \int_1^x dy\, D(y) \tag{50}$$

and differentiating the free energy with respect to $D(x)$, it is possible to derive the following integral equation for $D$:

$$\int_0^x dy \frac{D'(y)}{[yD(y) + \int_y dz D(z)]^3} = \frac{1}{4\pi^2 \rho a^3} \int_0^x dy\, D'(y) \int_0^\infty dp\, p^6 \frac{a(x)a(y)}{[\tilde{a} - \langle a \rangle - [a](y)]^2} \tag{51}$$

However, this equation is difficult to handle, even numerically, because for each $x$ it involves the value of $D(y)$ over the entire interval $y \in [0,1]$, making it impossible to grow the solution from $x = 0$. Therefore, instead of solving Eq. (51), we will show that it does not have a solution corresponding to an overlap parameter $q(x)$ that increases continuously from zero. This means that there exists a maximum value for the "tube diameter" $D_{\alpha\beta}^{1/2}$ (see Fig. 1 for illustration), and there is no such phenomenon as "freezing on all length scales." To proof this point, first note that the second term on the right hand side of Eq. (45) is zero, because according to Eq. (13) we have $K(0) = 0$, so $D(0) = \infty$, and $a(0) = 0$. By combining Eqs. (13), (42) and (45), one can write the expression for the mean-field free energy $F = \lim_{n \to 0} n^{-1} H_n^{\text{eff}}$ in the following way:

$$\frac{4F}{Na^2 \kappa_p^2 k_B T} = -\int_0^1 \frac{dx}{x^2} \tilde{K}(x) + \frac{2}{\pi^{3/2}} \frac{\ell^{1/2}}{\rho^{1/2} a^2} \int_0^\infty d\tilde{k}\, \tilde{k}^2 \left[ \ln\left(1 - \frac{\langle a \rangle}{1 + \tilde{k}^2}\right) - \int_0^1 \frac{dx}{x^2} \ln\left(1 - \frac{[a](x)}{1 + \tilde{k}^2 - \langle a \rangle}\right) \right]$$

$$a(x) = \exp\left[-\tfrac{1}{4} \tilde{D}(x) \tilde{k}^2\right] \qquad \tilde{k} = \kappa_p^{-1} k$$

$$\tilde{K}(x) = \kappa_p^{-2} K(x) \qquad \tilde{D}(x) = \kappa_p^2 D(x)$$

(52)

As a reminder we note that the first contribution to the free energy has an entropic origin, while the second contribution has an energetic origin. In order to arrive at this expression, the $Q_{\alpha\beta}$-independent term $\ln \tilde{a}$ has been subtracted from $E(k)/k_B T$, transforming the first term on the right hand side of Eq. (45) into $\ln(1 - \langle a \rangle / \tilde{a})$. Note that $\tilde{D}$ and $\tilde{K}$ are still related to each other via Eq. (14). Substituting Eq. (1) for the density it follows that the numerical factor in front of the energy is of order unity. As mentioned before, an important question is whether or not the sequence model, like the independent interaction model, permits a solution $\tilde{K}(x)$ that increases continuously from zero. Physically this would mean that there are replica pairs having an arbitrary small, but non-zero overlap. The existence of such a solution is suggested by the fact that the unscreened Coulomb interaction has no intrinsic length scale. Since $\tilde{D} = \kappa_p^2 D$ and $\tilde{K} = \kappa_p^{-2} K$ are rescaled quantities, overlap at a length scale which is large compared to the self-screening length corresponds to $\tilde{D} \gg 1$, and so $\tilde{K} \ll 1$. Since $q(x)$ has to be an increasing function[26] of $x$, the same must be true for $\tilde{K}(x)$. Also, according to Ref. 30 it is to be expected that there is an interval $[x_0, 1]$ of $x$-values where $\tilde{K}(x)$ is constant, and of order unity. Therefore, we look for a solution $\tilde{K}(x)$ that increases continuously from zero at the left side of the interval $[0,1]$, and reaches a plateau value $\tilde{K}_0 \approx 1$ at the right side (the solution Fig. 3 to the independent interaction model has these properties if $\kappa = 0$). The most obvious thing to do is to assume that for small $x$-values the function $\tilde{K}(x)$ scales like $\tilde{K}(x) \propto x^{m-1}$. We must have $m > 1$, and for $m < 3$ the dominant contribution to the integral representing the entropy comes from small $x$-values. Before inserting this type of solution into Eq. (52) for the free energy it is useful to make the following observations:

- For the replica symmetric solution $\tilde{K}(x) \equiv 0$ one has $a(x) \equiv 0$, and so both the entropy and the energy are zero. Therefore, in case of broken replica symmetry, the free energy must be positive.

- The entropy, as given in Eq. (13), is always negative, and so is the first contribution to the energy.

- The second contribution $E_2$ to the energy in Eq. (52) is the only positive term, and must at least counterbalance the entropy in case of replica symmetry breaking. Note that $E_2$ is smaller than the term $E_2'$ obtained by omitting the $\tilde{k}^2$-term from the denominator (Eq. (53)). Since this $\tilde{k}^2$-term originates from the potential, removing this term corresponds to considering an infinitely strong interaction.

$$E_2' \equiv -\int_0^1 \frac{dx}{x^2} \int_0^\infty d\tilde{k}\, \tilde{k}^2 \ln\left(1 - \frac{[a](x)}{1 - \langle a \rangle}\right) \qquad S = -\int_0^1 \frac{dx}{x^2}\, \tilde{K}(x) \qquad (53)$$

In the appendix it is shown that for functions $\tilde{K}(x)$ which scale like $x^{m-1}$ for small $x$, the entropy $S$ is dominant over $E_2'$, which implies that it is dominant over the energy $E$, and the free energy as given in Eq. (52) is negative. Since the replica-symmetric solution has free energy zero, solutions which increase continuously from zero following a power law cannot exist. We believe that it is justified to make the stronger statement that the maximization condition of the free energy given in Eq. (52) does not have a solution corresponding to an overlap parameter $q(x)$ that increases continuously from zero. Note that since in the appendix it is proven that $S$ is dominant over $E_2'$, our statement concerning the non-existence of such a solution applies to all situations for which $V(k) \geq 0 \ \forall k$. This condition is fulfilled if unlike particles attract each other ("mixing case"), and the (isotropic) interaction strength decays not slower than $1/r$ with the distance $r$. Although the exact solution to the maximization condition of the free energy may still be non-trivial, it follows from our analysis that the value of $\tilde{K}(x)$ must be of order unity wherever it is non-zero. Therefore, it is not too bad an approximation to consider a 1-step solution, i.e.

$$\begin{aligned} \tilde{K}(x) &= 0 & x \leq x_0 \\ \tilde{K}(x) &= K_0 & x > x_0 \end{aligned} \qquad (54)$$

Investigating this solution numerically shows that the factor in front of the energy (see Eq. (52)) should at least equal 8 in order to make the solution stable. It follows that in order to get replica symmetry breaking the density should be much lower than the one given in Eq. (1). If the chain is present in a poor solvent, the situation is even worse; see the discussion following Eq. (49). Therefore, we believe that it is safe to conclude that there is no replica symmetry breaking under these conditions. In the summary we discuss under what conditions replica symmetry breaking *can* occur, and draw some qualitative conclusions from the results obtained in the present section.

## VI. SUMMARY

In this paper we investigated the possibility of freezing in the globular state of a "statistically neutral" random polyampholyte chain in a $\theta$-solvent for the situation where the polymer self-screening length is much larger than the average distance between two charged monomers (weakly charged case). Under these conditions the globule is far from being maximally compact and still contains a lot of solvent. Due to some general considerations (see the introduction) it was to be expected that the long-range character of the interaction changes the nature of the freezing transition. Using the analytical tools developed in Ref. 15, we first studied the globule in the independent interaction approximation. This lead to the prediction that if the salt screening length $\kappa^{-1}$ is larger than the polymer self-screening length $\kappa_p^{-1}$, the replica symmetry breaking is non-trivial on length scales $R$ satisfying $\kappa_p^{-1} < R < \kappa^{-1}$. However, one should be cautious in accepting the predictions made using the independent interaction model, because in this model self-screening does not exist. Therefore, we checked the results using the more realistic (and more complicated) sequence model. It was shown that on length scales smaller than $\kappa_p^{-1}$ the sequence model is equivalent to the independent interaction model, with the important footnote that the parameter $\kappa^{-1}$, which represents the salt screening length in the independent interaction model, now represents the self-screening length $\kappa_p^{-1}$. Adopting this physical interpretation of $\kappa^{-1}$, the interesting regime $\kappa_p^{-1} < R < \kappa^{-1}$ found in Section IV does not exist. In order to study the globule at length scales $R \gtrsim \kappa_p^{-1}$ one has to resort to the sequence model. We showed that for any interaction that favors mixing of unlike particles, and which decays not slower than $1/r$, the overlap parameter $q(x)$ is discontinuous in the sense that it must be of order unity wherever it is non-zero. Physically, this can be attributed to self-screening, which renders the interaction effectively short-range. For weakly charged polyampholytes in a $\theta$-solvent or a good solvent we arrived at the conclusion that although the system is frustrated to a certain extent, no replica symmetry breaking occurs. In a poor solvent, the situation is even more relaxed: the frustration due to the presence of the polymer bonds is less if the monomers are confined to a smaller volume. At the same time, this suggests a way to increase the frustration in order to attain a state with replica symmetry breaking: by forcing the chain to remain in a state with *smaller* density than the one given by Eq. (1). In such an extended state the constraints imposed by the polymer bonds are more important, and the frustration is larger. This situation can be achieved in the following way. If the faction $f$ of charged monomers increases, the density becomes larger. For a certain value of $f$, the density reaches its maximum value $\rho_{max} = f/b^3$ corresponding to the compact state. If $f$ increases beyond this value, the electrostatic attractions become stronger, but the chain is not able to respond by further collapse. So, it has to stay in the "extended" state corresponding to the maximum density. Of course, Eq. (1) breaks down before this point is reached. Also, in order to obtain an estimate for the fraction of charged monomers needed to attain the compact state, one should take into account the fact that in a dense globule the value of the Bjerrum length $\ell$ is much larger than its value $\ell \approx 0.7 nm$ in pure water. Summarizing, the only possibility to attain replica symmetry breaking in polyampholytes is by considering maximally compact globules with a large fraction of charged monomers. However, in this case the screening length is of the same order of magnitude as the monomer size $b$, and the analytical analysis presented in this article breaks down. Nevertheless, we believe that the results of Section V remain qualitatively valid in the sense that the freezing transition can at least qualitatively be described by a 1-step replica symmetry breaking,[24] and the system behaves in the same way as a globule with short-range interactions. Since we attribute this to the presence of self-screening, this conclusion is probably only true in the "mixing case", where unlike particles attract each other. In the segregating case, screening does not exist and the long-range character will have an important effect on the nature of the freezing transition.

## ACKNOWLEDGEMENTS

H. J. Angerman is grateful to the NWO (Dutch Organization for Scientific Research) for financial support in the form of a NATO-Science Fellowship.

# APPENDIX A

In this appendix we proof that if the function $K(\tilde{x})$ follows a power law $\tilde{K}(x) \propto x^{m-1}$ for small $x$, then the entropy $S$ is dominant over the energy $E_2'$ (the definitions for $S$ and $E_2'$ are reproduced in equation (A1)). As explained in the main text, the implication of this result is that in polymer globules with two kinds of monomers where unlike monomers attract each other, the overlap parameter $q(x)$ must be of order unity wherever it is non-zero, even if the interaction is long-range (sufficient condition: $V(k) > 0 \ \forall k$). This means that there exists a maximum value for the "tube diameter" (see Fig. 1).

$$E_2' = -\int_0^1 \frac{dx}{x^2} E(x) \qquad S = -\int_0^1 \frac{dx}{x^2} \tilde{K}(x)$$

(A1)

$$E(x) \equiv \int_0^\infty d\tilde{k}\ \tilde{k}^2 \ln\left(1 - \frac{[a](x)}{1-\langle a \rangle}\right)$$

We proceed as follows. For $\tilde{K}(x) \propto x^{m-1}$ we show that $E(x) \propto x^{m'-1}$ for small $x$, and find $m'$ in terms of $m$. In order for the (positive) energy to be dominant over the (negative) entropy, $m'$ must be smaller than $m$. We will show that this condition cannot be satisfied. Using the expressions for $\langle a \rangle$ and for $[a](x)$ given in Eqs. (45) and (46), the expression for $E(k)$ becomes (the tilde on the integration variable $\tilde{k}$ will be omitted, and the prefactor $c$ in $\tilde{D}(x) = cx^{-m}$ is absorbed in $k$)

$$E(x) = \int_0^\infty dk\ k^2 \ln\left(1 - \frac{x\exp[-x^{-n}k^2] - n^{-1}k^{2/n}\Gamma[-n^{-1}, x^{-n}k^2]}{1 - n^{-1}k^{2/n}\ \Gamma[-n^{-1}, k^2]}\right)$$

$$\Gamma(s,x) \equiv \int_x^\infty dy\ y^{s-1} \exp[-y]$$

(A2)

In order to find an upper bound for $E(x)$, we construct a lower bound $f(k)$ for the denominator in Eq. (A2); that is, we construct a function $f(k)$ such that

$$f(k) \leq g(k) \equiv 1 - n^{-1}k^{2/n}\Gamma[-n^{-1}, k^2] \qquad \forall\ k \geq 0 \qquad \text{(A3)}$$

The dependence of $g(k)$ on $k$ is shown in Fig. 5 for $n = 2$, in which case it starts at zero with finite slope, and converges to unity for large $k$. For general values of $n$, the first term of the series expansion is

$$g(k) = -n^{-1}\Gamma[-n^{-1}] k^{2/n} + \cdots \tag{A4}$$

Note that $\Gamma[-n^{-1}]$ is negative. We split the interval $0 \leq k \leq \infty$ in two parts, separated by $k_* = x^{n/2}$. In the left part we choose $f(k)$ to be given by the first term of the series expansion of $g(k)$, in the right part we choose $f(k)$ to be constant. More precisely (see Fig. 5 for illustration for the case $n = 2$ for which $f(k)$ is linear on $0 \leq k \leq k_*$).

$$\begin{cases} f(k) = -r n^{-1} \Gamma(-n^{-1}) k^{2/n} & k < x^{n/2} \\ f(k) = -r n^{-1} \Gamma(-n^{-1}) x & k \geq x^{n/2} \end{cases}$$

$$r = 1 - \varepsilon \qquad \varepsilon \ll 1 \tag{A5}$$

Since the numerator in Eq. (A2) is a decreasing function of $k$ for $k > x^{n/2}$, the argument of the logarithm will not become negative. Substituting $f(k)$ for $g(k)$ in Eq. (A2) leads to the sum of two integrals. In both integrals, the integrand is a function of $p \equiv k\, x^{-n/2}$, and we obtain

$$|E(x)| < x^{3n/2} \int_0^1 dp\, p^2 \ln\left(1 - \frac{p^{-2/n}\exp[-p^2] - n^{-1}\Gamma[-n^{-1}, p^2]}{-r n^{-1}\Gamma[-n^{-1}]}\right) +$$

$$+ x^{3n/2} \int_1^\infty dp\, p^2 \ln\left(1 - \frac{\exp[-p^2] - n^{-1} p^{2/n}\Gamma[-n^{-1}, p^2]}{-r n^{-1}\Gamma[-n^{-1}]}\right) \tag{A6}$$

For given value of $n$, the integrals in equation 1 are just numerical factors of order unity. Both terms in Eq. (A6) scale as $x^{3n/2}$. Since $\tilde{K}(x) \propto x^{n-1}$, the negative entropy is dominant over the positive energy unless $n \leq -2$, which is impossible since $\tilde{K}(x)$ has to be an increasing function of $x$.

## Legends to the Figures

FIG. 1. Pictorial way to describe the trial function Eq. (12) for the overlap parameter. The conformations of replicas $\alpha$ and $\beta$ coincide when the monomer positions are measured with accuracy $D_{\alpha\beta}^{1/2}$. When observed at better resolution, the conformations turn out to be different.

FIG. 2. Qualitative representation of the non-trivial solution in Eq. (27). The rescaled quantities $x$ and $K$ are defined in Eq. (28). The solution has two branches. The dotted branch must be rejected since it leads to an overlap parameter $q(x)$ that decreases with increasing $x$. The solid branch represents the physical solution; it converges to the asymptot $K = x$ (dashed line).

FIG. 3. Approximate shape of $K(x)$ calculated within the independent interaction approximation for the situation $\kappa^{-1} \gg \kappa_p^{-1}$. The jump in $K$ is due to the absence of freezing at length scales larger than $\kappa^{-1}$. The interval where $K$ is linear corresponds to length scales in between $\kappa_p^{-1}$ and $\kappa^{-1}$; in this range the freezing is qualitatively different from that occurring in polymer globules with short-range interactions. The plateau on the right corresponds to freezing on the length scale $\kappa_p^{-1}$. If the independent interaction model is obtained from the sequence model by expansion of the free energy in powers of the potential, then $\kappa = \kappa_p$, and the solution shown here does not exist for weakly charged polyampholytes.

FIG. 4. This figure shows, for the independent interaction model, the probability density $P$ that the overlap between two randomly picked replicas equals $q$. The delta peak at the left represents the non-zero probability that two replicas have no overlap at all. The delta peak at the right represents the probability that two replicas have perfect overlap down to the self-screening length scale. The probability to find a larger overlap is zero, because at length scales smaller than the self-screening length the thermal energy prevents freezing. The continuous part of $P(q)$ represents non-trivial freezing due to the long-range character of the interaction (compare with Ref. 15).

FIG. 5. The solid curve represents the denominator $g(k) \equiv (1 - n^{-1} k^{2/n} \Gamma[-n^{-1}, k^2])$ of the quotient in Eq. (A2) for the case $n = 2$; the dashed curve represents its lower bound $f(k)$ used to find an upper bound for the energy $E(x)$. For $n \neq 2$ the dashed curve will not be linear for small $k$ but increase according to the power law $f(k) \propto k^{2/n}$.

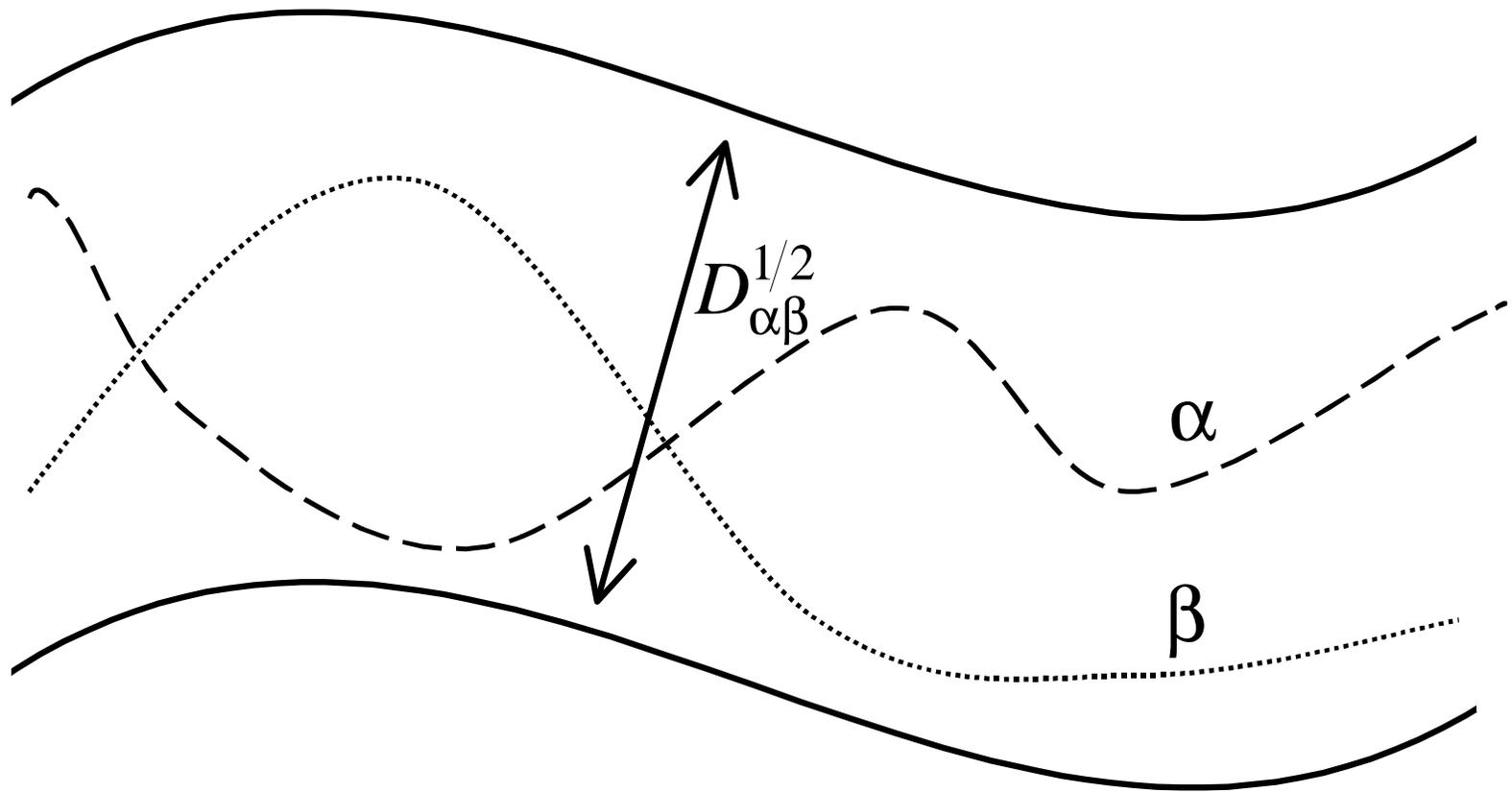

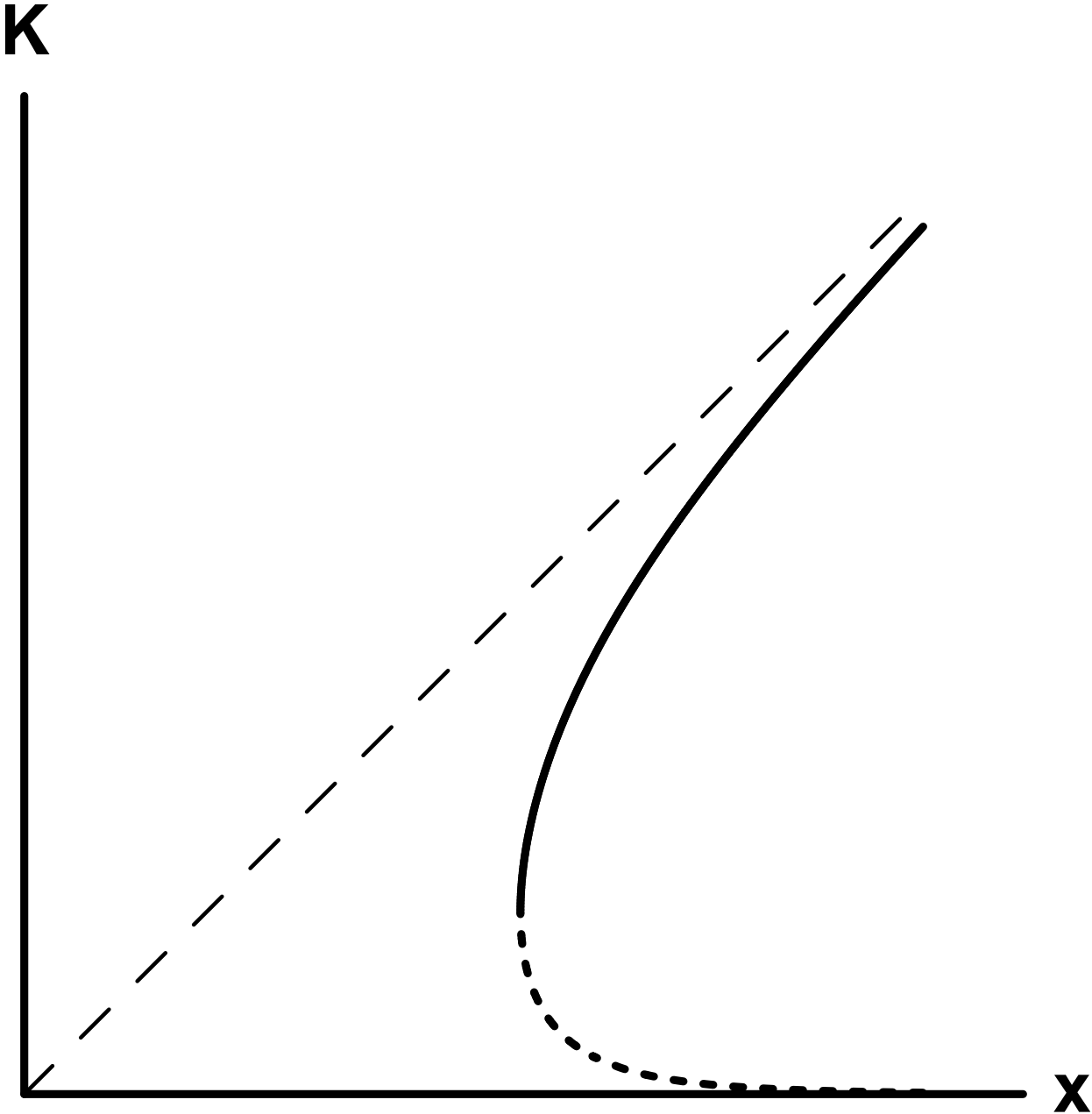

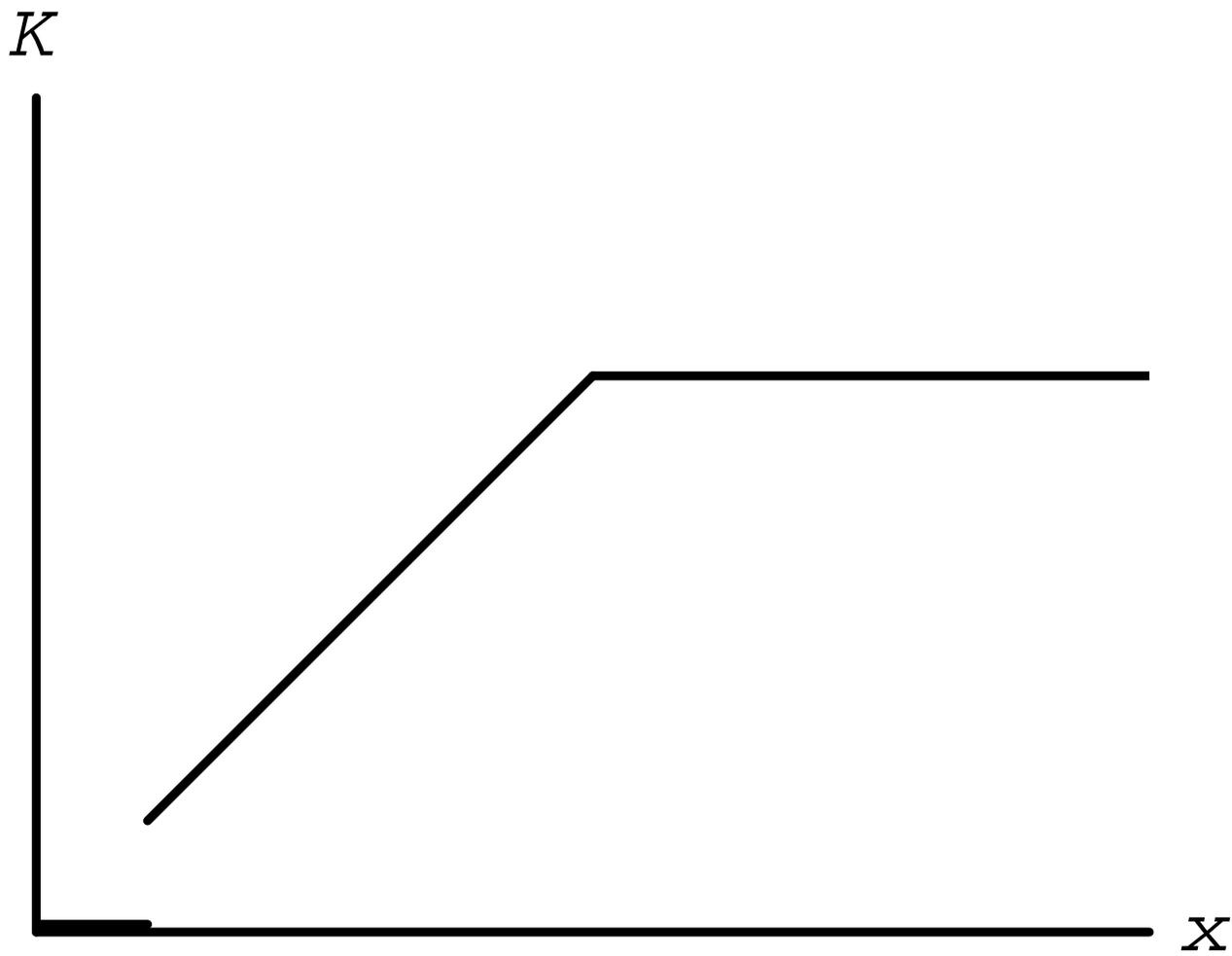

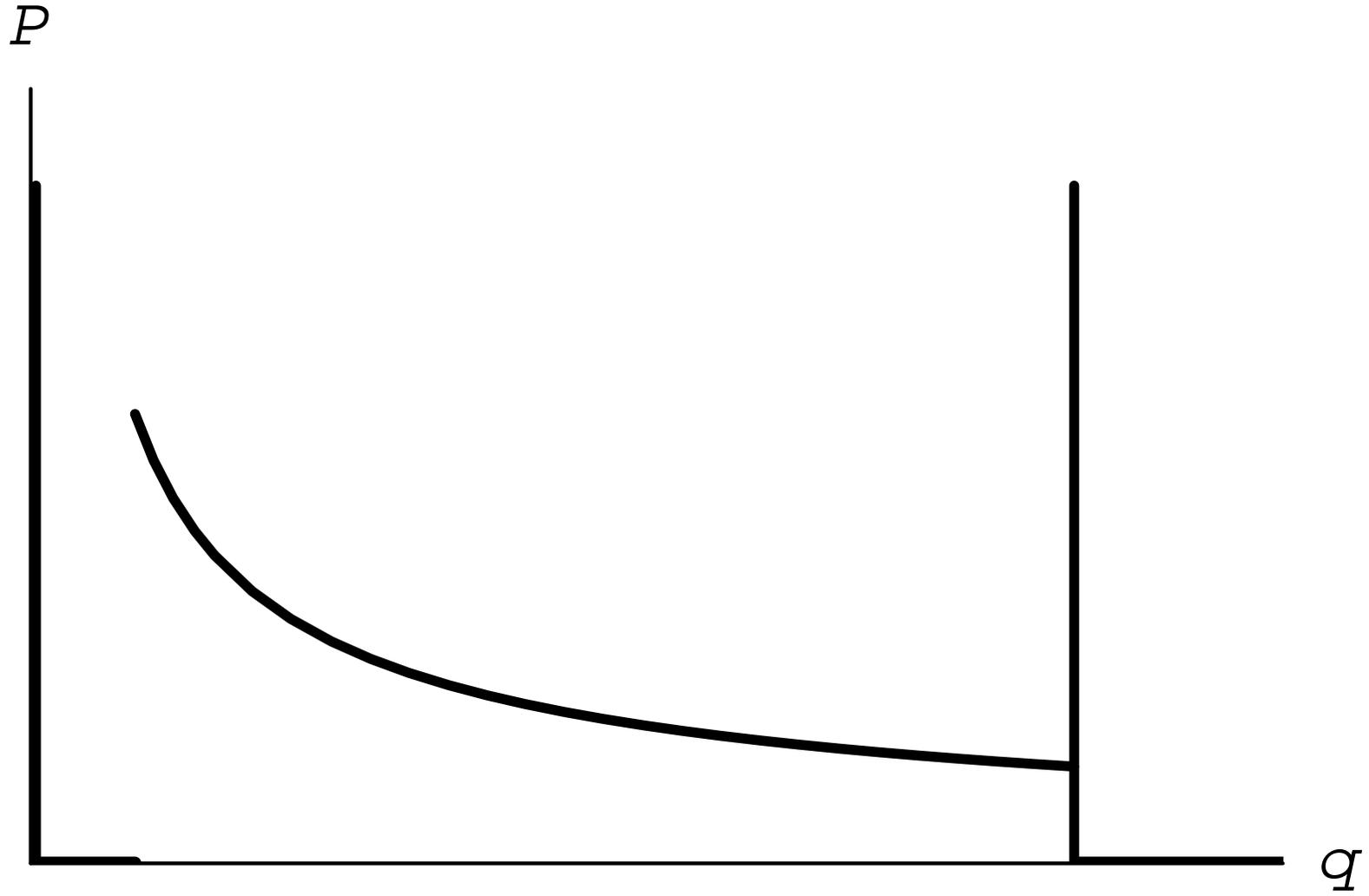

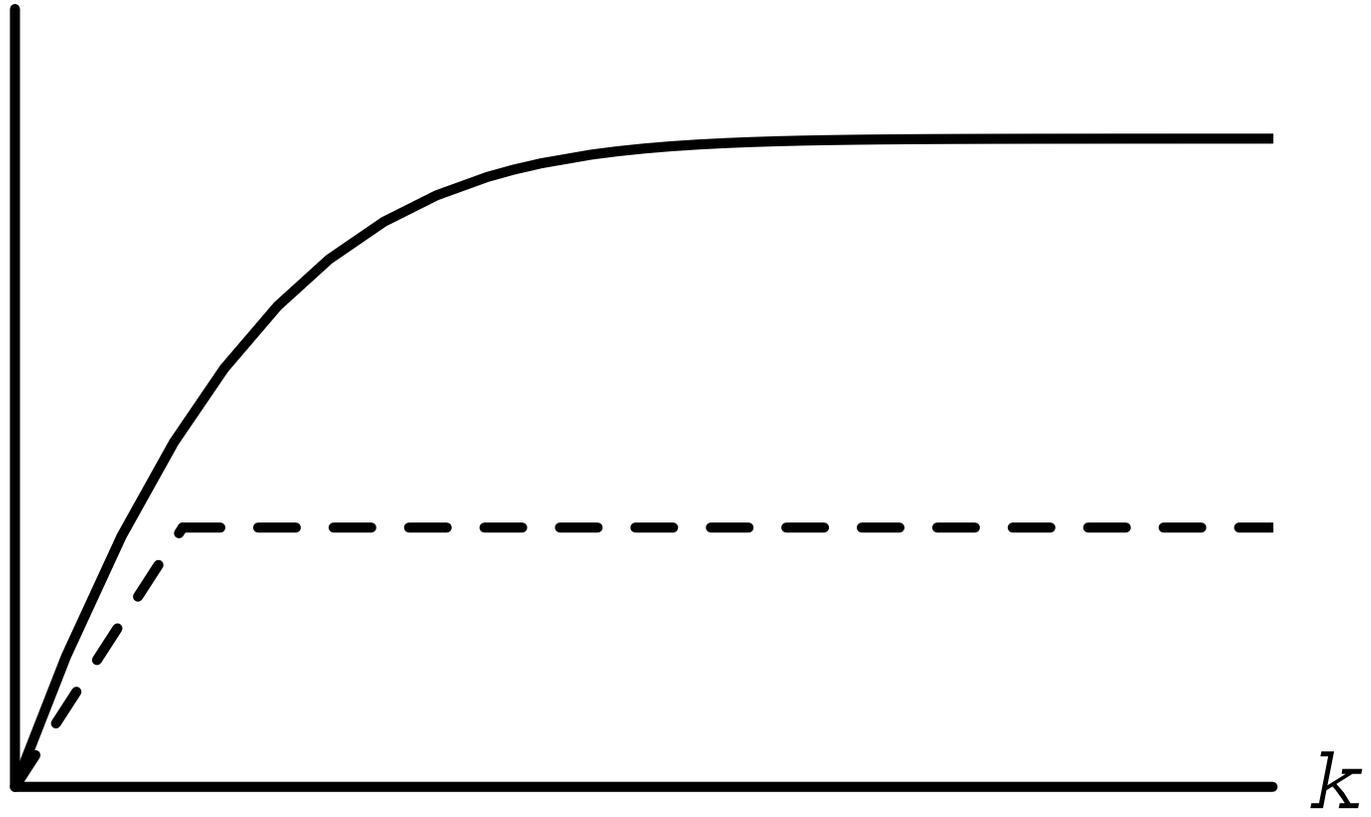